\documentclass[ChapterTOCs,krantz1]{krantz}

\usepackage{amssymb,amsmath,bbm}
\usepackage{graphicx,subfigure}
\usepackage{makeidx}
\usepackage{multicol}
\usepackage{epsfig,color}

\usepackage[sectionbib]{bibunits}
\newcommand{\be}{\begin{eqnarray}}
\newcommand{\ee}{\end{eqnarray}}
\newcommand{\bea}{\begin{eqnarray}}
\newcommand{\eea}{\end{eqnarray}}
\newcommand{\bma}{\begin{subequations}}
\newcommand{\ema}{\end{subequations}}

\DeclareMathOperator{\tr}{Tr} 



\makeindex

\begin{document}
\bibliographyunit[\chapter]
\defaultbibliography{biblio}
\defaultbibliographystyle{plain}

\title{Developments in Quantum Phase Transitions}



\frontmatter

\tableofcontents

\mainmatter

\setcounter{page}{1}
\chapterauthor{Sean A. Hartnoll}{Harvard University, Department of Physics, Jefferson Laboratory, Cambridge, MA 02138, U.S.A.}

\chapter{Quantum Critical Dynamics from Black Holes}

\section{Brief motivation}

Free or weakly interacting field theories underpin a great deal of our physical intuition. The fact that effective weakly interacting quasiparticles can emerge even when the `fundamental' microscopic degrees of freedom are strongly interacting, as illustrated dramatically by Landau's theory of Fermi liquids \cite{SAH:anderson}, has allowed great progress in physics without confronting head-on the problem of strong coupling. Even where genuinely strongly 
interacting physics is required, for instance at the Wilson-Fisher fixed point in 2+1 dimensions, computational techniques such as the (vector) large $N$ or $4-\epsilon$ expansions allow us to move the problem into a weakly coupled regime. At small $\epsilon$ or large $N$ one can compute, using weak coupling notions such as single-particle propagators, and then extrapolate back to the physical values of $\epsilon=1$ or $N=2,3$. Remarkably, this procedure often gives qualitative and even quantitatively correct answers. Such expansions are difficult and less accurate, however, for real time correlators at finite temperature (e.g. \cite{SAH:sachdevbook} or Sachdev's chapter in this book).

A weak coupling intuition also informs our expectations of the low temperature states of matter. Broadly speaking, in the presence of bosonic quasiparticles one expects a condensate, while fermions are expected to build up a Fermi surface.

One might ask whether there may be inherently strongly coupled physical phenomena and novel states of matter at low temperature that cannot easily be conceptualised from weak coupling. This question has become more pressing following the discovery of a plethora of physical systems exhibiting `non-Fermi liquid' behaviour, most infamously, the strange metal region of the cuprate superconductors (e.g. \cite{SAH:hightc}).

The holographic correspondence to be reviewed in this chapter presents model quantum field theories in which controlled computations may be performed directly at strong coupling. It is hoped that this approach may at the very least help to cast away some of the conceptual baggage of weak coupling. More ambitiously, by turning on a finite chemical potential in these models, one can hope to discover novel states of matter that could inform our theoretical approach to real world systems. For instance, we will encounter below cases of gapless charged bosons that do not condense at low temperatures, as well as a strongly coupled onset of superconductivity without quasiparticles. We will also uncover computationally controlled non-Fermi liquid behaviour in fermionic spectral densities, and indications of an underlying strongly coupled `Fermi surface'. Throughout, I will emphasise certain notions that arise, such as the importance of `quasinormal poles', that may inform general systematic approaches to strongly interacting matter.

\section{The holographic correspondence as a tool}

It has been argued that theories of quantum gravity obey a holographic principle (see e.g. \cite{SAH:Susskind}
for a pedagogical exposition of these ideas). This is the statement that the number of `local' degrees of freedom in gravitational theories should scale like the area enclosing a volume, in Planck units, rather than the volume itself. While this principle remains to be fully understood in general, following the seminal papers \cite{SAH:Maldacena:1997re, SAH:Witten:1998qj, SAH:Gubser:1998bc} it has been made precise in a specific framework which we will call the holographic correspondence.

There is a large amount of evidence that certain quantum field theories are holographically equivalent to theories of quantum gravity in one higher dimension (\cite{SAH:Horowitz:2006ct} and \cite{SAH:McGreevy:2009xe} are conceptual overviews with references)
\be
\begin{array}{c}
\text{Quantum field theory} \\
d \; \text{spacetime dimensions}
\end{array}
\quad \leftrightsquigarrow \quad
\begin{array}{c}
\text{Quantum gravitational theory} \\
d+1 \; \text{spacetime dimensions.}
\end{array}\label{SAH:eq:quantum}
\ee
The most established examples are ${\mathcal{N}}=4$ super Yang-Mills theory in 3+1 dimensions and the infrared fixed point of ${\mathcal{N}}=8$ super Yang-Mills theory in 2+1 dimensions. These are supersymmetric $SU(N)$ gauge theories. Their field content is a gauge field $A$ together with multiple scalar ($\Phi$) and fermionic ($\Psi$) fields transforming in the adjoint representation of $SU(N)$. All the fields can therefore be written as $N \times N$ matrices. The Lagrangian is schematically
\be\label{SAH:eq:CFTaction}
{\mathcal{L}}_\text{QFT} \sim \tr \left(F^2 + \left(\partial \Phi \right)^2 + i \bar \Psi \Gamma \cdot \partial \Psi + g^2 [\Phi,\Phi]^2 + i g \bar \Psi [\Phi, \Psi] \right) \,,
\ee
here $F = dA + g A\wedge A$ is the nonabelian field strength. There are no mass terms; in the 3+1 dimensional case the Yang-Mills coupling $g$ is exactly marginal, and the theory is conformal at all couplings. In 2+1 dimensions, the coupling runs to a strongly coupled infrared fixed point.

The right hand side of the correspondence (\ref{SAH:eq:quantum}) has still not been completely defined and is difficult to work with. What makes the correspondence useful is a simplification that occurs in the t'Hooft large $N$ limit \cite{SAH:'tHooft:1973jz, SAH:coleman} of these gauge theories. In this limit the gravitational theory becomes classical
\be
\begin{array}{c}
\text{Large $N$ gauge theory} \\
d \; \text{spacetime dimensions}
\end{array}
\quad \leftrightsquigarrow \quad
\begin{array}{c}
\text{(Semi)classical gravitational theory} \\
d+1 \; \text{spacetime dimensions.}
\end{array}
\ee
The essential feature of the t'Hooft limit that allows a dual classical description is that it induces a `large $N$ factorisation' of single trace operators. Namely, if the operator ${\mathcal{O}}(x) = \tr \left( \cdots \right)$ then the disconnected part of $n$-point functions are bigger by powers of $N$ than connected contributions. Thus the single trace operators are classical, or `mean field', for many purposes. For instance
\be
\langle{\mathcal{O}}(x) {\mathcal{O}}(y) \rangle = \langle{\mathcal{O}(x)} \rangle \langle {\mathcal{O}}(y) \rangle + {\mathcal{O}}\left( N^{-2} \right) \,.
\ee
However, unlike in the large $N$ limit of vector models, the theory remains strongly coupled. For instance, the operators 
${\mathcal{O}}(x)$ generically have large anomalous dimensions and the field theory will not be describable in terms of quasiparticles. The t'Hooft limit of gauge theories retains more features of the strongly coupled theories of physical interest. 

While an extra dimension of spacetime may seem mysterious to the field theorist, it has a clear physical meaning: it is the renormalisation group scale. The renormalisation group flow equations are differential equations for couplings that are local in the energy scale. The holographic correspondence realises this locality on an equal footing with spacetime locality of the field theory. Certain components of the gravitational field equations, determing the evolution of the `bulk' spacetime along the extra dimension, will precisely correspond to the renormalisation group equations of the `boundary' field theory. The picture that one has in mind is the following figure \ref{SAH:UVIR}:

\begin{figure}[h]
\begin{center}
\includegraphics[height=120pt]{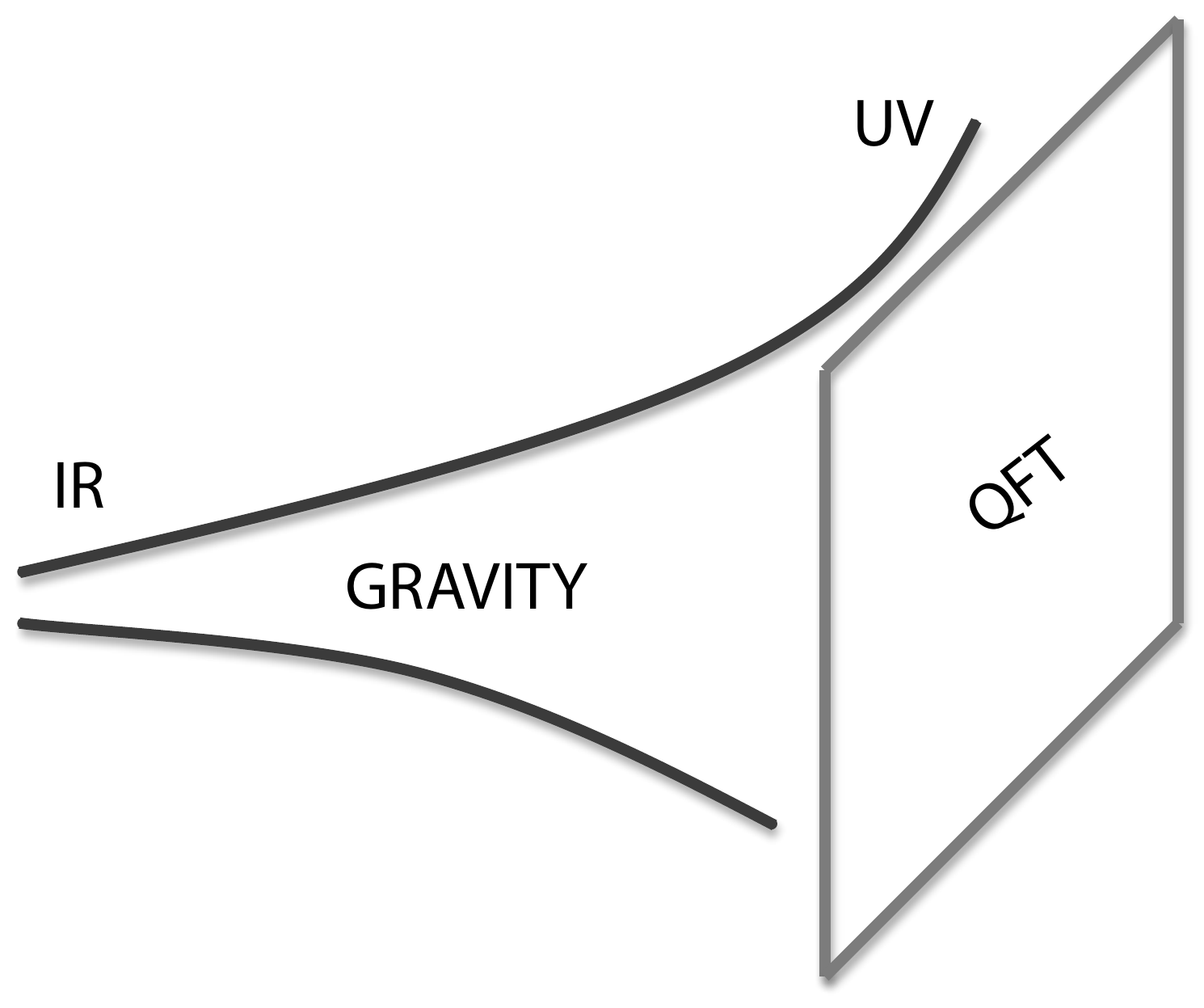}
\end{center}
\caption[Energy scales in the holographic correspondence.]{The extra spatial dimension of the gravitational dual represents
the renormalisation group flow of the quantum field theory.}\label{SAH:UVIR}
\end{figure}

The first ingredient needed to make the correspondence more precise is an action for the bulk theory. This will underlie the dual description of the field theory given by (\ref{SAH:eq:CFTaction}).
The classical gravitational action in $d+1$ dimensions will depend on the dual $d$ dimensional large $N$ field theory.
The full action will generically be complicated, with infinitely many fields. However, often it is possible to consistently truncate the full action to a small number of fields that capture the physics of interest. We can start with the most universal sector of the theory, which is the action for the metric $g_{ab}$. The simplest action we can consider is the Einstein-Hilbert action together with a negative cosmological constant:
\be\label{SAH:eq:einstein}
S[g] = \frac{1}{2 \kappa^2} \int d^{d+1}x \sqrt{-g} \left(R + \frac{d (d-1)}{L^2} \right) \,.
\ee
Here $R$ is the Ricci scalar, $L$ is a lengthscale whose meaning will become apparent shortly and $\kappa$ is the gravitational constant. A modern introduction to general relativity is \cite{SAH:carroll}. The Einstein equations of motion following from this action are
\be\label{SAH:eq:eeqs}
R_{ab} = - \frac{d}{L^2} g_{ab} \,,
\ee
where $R_{ab}$ is the Ricci tensor. The most symmetric solution to these equations is
Anti-de Sitter space (AdS) which has the metric
\be\label{SAH:eq:ads}
ds^2 = L^2 \left(\frac{-dt^2 + dx^i dx^i}{r^2} + \frac{dr^2}{r^2} \right) \,.
\ee
We think of the coordinates $\{t,x^i\}$ as parametrising the space on which the field theory lives,
while $r$ is the extra radial coordinate in figure \ref{SAH:UVIR} running from $r=0$ (the `boundary') to $r=\infty$ (the `horizon'). Two aspects of this solution should be discussed: its symmetries and the overall scale $L$.

The full isometry group of the spacetime (\ref{SAH:eq:ads}) is $SO(d,2)$. This is precisely the conformal group in $d$ dimensions. The symmetries of the bulk action act on the `boundary' quantum field theory (QFT) as conformal transformations \cite{SAH:Witten:1998qj, SAH:Gubser:1998bc}, indicating that the dual QFT is conformally invariant. In particular, the scaling symmetry of the QFT acts on the spacetime as
\be\label{SAH:eq:scaling}
\{ t, x^i, r \} \to \{\lambda t, \lambda x^i, \lambda r\} \,, 
\ee
which clearly leaves the metric (\ref{SAH:eq:ads}) invariant. This gives a first indication of why the radial direction $r$ is associated with the energy scale: as we scale to late times, we also scale towards large values of $r$, consistent with
these being the low energy, or `IR' region of figure \ref{SAH:UVIR}.

The overall factor of $L$ in (\ref{SAH:eq:ads}) sets the radius of curvature of the AdS spacetime. In order for classical gravity to be a valid description we need this radius to be large in Planck units. This requires
\be\label{SAH:eq:limit}
c \sim \frac{L^{d-1}}{\kappa^2} \gg 1 \,.
\ee
This limit ensures that quantum corrections to (\ref{SAH:eq:einstein}) generated by graviton loops will be small,
allowing us to focus on classical solutions to the Einstein equations. We denoted the ratio in (\ref{SAH:eq:limit}) by $c$ because it is the area enclosing a spatial volume of AdS in Planck units. The holographic principle suggests that this quantity should be associated with the number of degrees of freedom of the dual field theory. For more precise arguments see \cite{SAH:Susskind:1998dq,SAH:Susskind,SAH:McGreevy:2009xe}. Indeed, we will see below that $c$ is proportional to the central charge of the QFT defined, for instance, as the coefficient of the free energy when the theory is placed at a high temperature (see e.g. \cite{SAH:Kovtun:2008kw}). In most of the established examples of the correspondence for $3+1$ dimensional gauge theories, $c \sim N^2$, where $N$ is the rank of the gauge group of the QFT with action (\ref{SAH:eq:CFTaction}). We will therefore refer to (\ref{SAH:eq:limit}) as the `large $N$' limit. In general $c$ will scale like some positive power of $N$. If we were to latticize the QFT, $c$ would indicate the number of degrees of freedom per site.

Two further comments are necessary regarding the limit (\ref{SAH:eq:limit}) of the action (\ref{SAH:eq:einstein}). The first is that it is an unnatural tuning from a Wilsonian point of view, effectively ignoring the cosmological constant problem by imposing that the cosmological constant be small in Planck units. One might worry whether there are any theories of quantum gravity for which this is a consistent limit to impose. Secondly, there could in principle be higher derivative terms, such as curvature squared terms $R^2$, in the classical action. These terms can and have been incorporated into the holographic correspondence, but they increase the difficulty of computations and introduce additional parameters. If the couplings of such terms are natural, then they will be highly suppressed in the limit (\ref{SAH:eq:limit}). For peace of mind, we would like to know if there are concrete examples of gravitational theories which are described by the Einstein-Hilbert action (\ref{SAH:eq:einstein}) together with the limit (\ref{SAH:eq:limit}).

The large $N$ limit of the best studied examples of the holographic correspondence, ${\mathcal{N}}=4$ super Yang-Mills theory in 3+1 dimensions and ${\mathcal{N}}=8$ super Yang-Mills theory in 2+1 dimensions, are dual to so-called `Freund-Rubin' compactifications of string theory which do indeed have sectors described by the Einstein-Hilbert action with a weakly curved AdS vacuum. However, there are many more theories for which this is the case. These are found in the `landscape' of string theory vacua (see e.g. \cite{SAH:Denef:2008wq}). The string landscape was in fact discovered in attempting to justify fine tuning the cosmological constant \cite{SAH:Bousso:2000xa}. In the present context the landscape indicates the existence of many conformal field theories with large central charge and a universal sector dually described by the Einstein-Hilbert action \cite{SAH:Denef:2009tp}. More generally, the existence of a vast landscape of string vacua enables us to fine tune the cosmological constant and then use effective field reasoning in the bulk gravitational description. While it may be that some effective field theories cannot be completed into a UV consistent theory of quantum gravity (such theories are sometimes said to belong to the `swampland' \cite{SAH:Vafa:2005ui}), the existence of a large number of effective field theories that can be UV completed (the string landscape) will allow us to ignore considerations of quantum gravity UV consistency for the remainder.

Granted the existence of many QFTs with classical gravity duals, we still do not have a precise characterisation of this set of QFTs. Having a large central charge, as in (\ref{SAH:eq:limit}), is one necessary feature. Another appears to be that there is a parametrically large gap in the spectrum of anomalous dimensions of operators in the QFT \cite{SAH:Heemskerk:2009pn}. This means that there are only a handful of relevant operators, with most operators being highly irrelevant. While all of the well understood examples of theories with holographic duals are supersymmetric, it is not clear at the time of writing whether this is an essential ingredient or a technical crutch. Finally, while I will focus in this chapter on quantum critical theories that are Lorentz invariant, the holographic correspondence has recently been extended to theories with dynamical critical exponent $z \neq 1$ \cite{SAH:Son:2008ye, SAH:Balasubramanian:2008dm, SAH:Kachru:2008yh}. I will furthermore set the speed of light equal to unity throughout.

\subsection{The basic dictionary}

The most basic and useful entry in the holographic correspondence dictionary is that
for every gauge invariant operator ${\mathcal{O}}$ in the QFT, there is a corresponding field $\phi$ in
the bulk gravitational theory:
\be
\begin{array}{c}
\text{operator ${\mathcal{O}}$} \\
\text{(quantum field theory)}
\end{array}
\quad \leftrightsquigarrow \quad
\begin{array}{c}
\text{dynamical field $\phi$} \\
\text{(bulk)\,.}
\end{array}
\ee
So far the only bulk field we have discussed is the metric $g_{ab}$. By the above statement, this should
be dual to an operator in the QFT, but which one? Given that the metric is universal, present by definition
in all classical theories of gravity, we should expect to match the metric to an operator that exists in
all QFTs. Furthermore, we can expect the operator to have spin 2, just like the graviton. The natural
guess is that the correct operator is the energy momentum tensor $T^{\mu \nu}$ of the QFT. The indices
$\mu,\nu$ run over the $d$ spacetime dimensions of the QFT while $a,b$ run over the $d+1$ dimensions
of the bulk. In many situations we can work in a gauge in the bulk in which fluctuations of the metric
(gravitons) have no components in the radial direction, $\delta g_{ar} = 0$, thus allowing a matching between
components of metric fluctuations and components of the QFT energy momentum tensor.

\begin{figure}[h]
\begin{center}
\includegraphics[height=125pt]{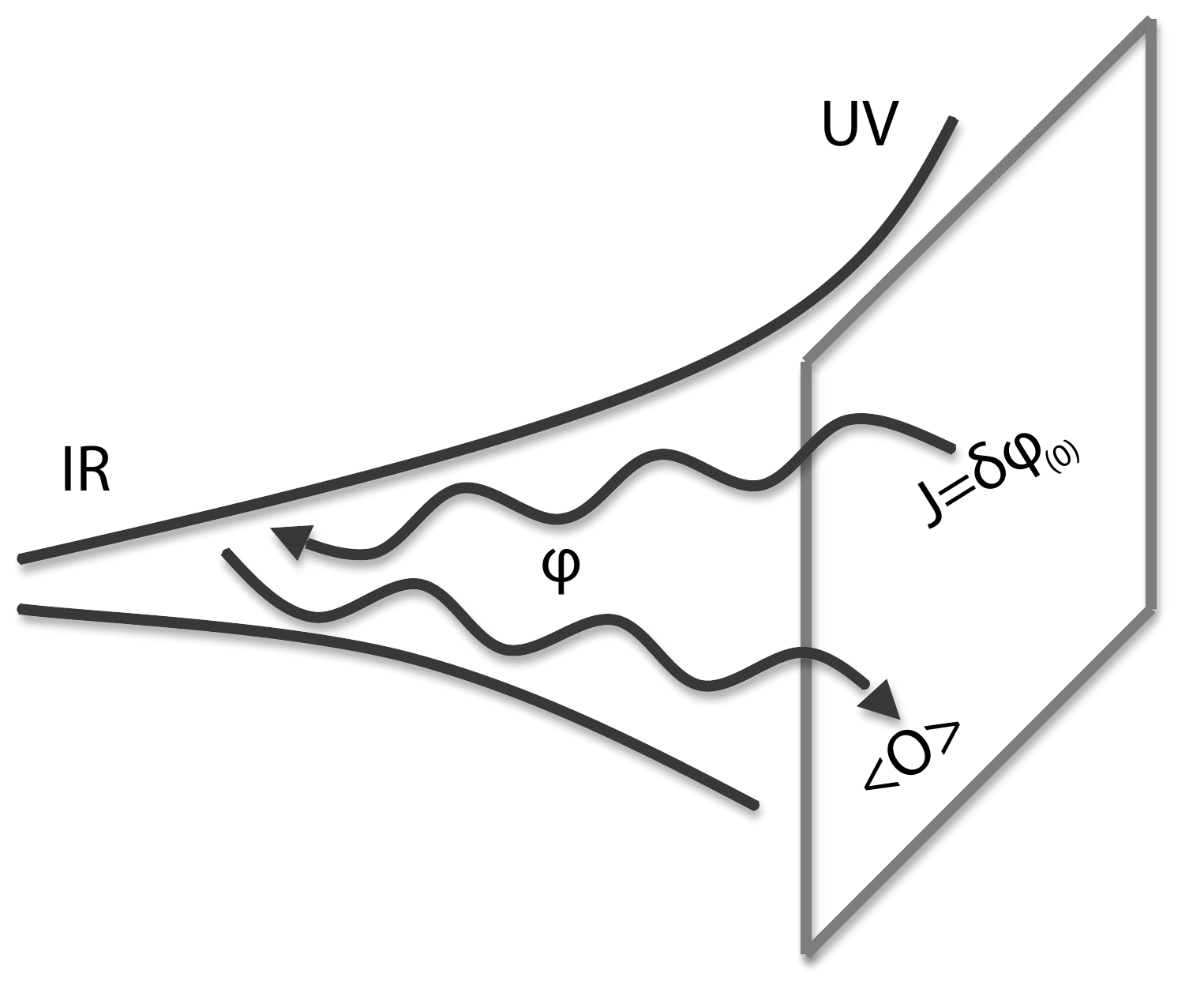}
\end{center}
\caption[Sources are dual to boundary values.]{A source $J$ for an operator ${\mathcal{O}}$ in the QFT
corresponds to a boundary condition $\delta \phi_{(0)}$ for the bulk field $\phi$ dual to ${\mathcal{O}}$. Solving the bulk equations of motion for $\phi$ allows computation of the expectation value $\langle {\mathcal{O}} \rangle$ due to the source $J$.}\label{SAH:response}
\end{figure}

More generally the bulk action will contain more fields than just the metric. Each field in the bulk will correspond to
an additional operator in the QFT. Examples of the field-operator correspondence we will need here are:
\be\label{SAH:eq:list}
\begin{array}{c}
\text{energy momentum tensor: $T^{\mu \nu}$} \\
\text{global current: $J^{\mu}$} \\
\text{scalar operator: ${\mathcal{O}}_B$} \\
\text{fermionic operator: ${\mathcal{O}}_F$}
\end{array}
\quad \leftrightsquigarrow \quad
\begin{array}{c}
\text{graviton: $g_{ab}$} \\
\text{Maxwell field: $A_a$} \\
\text{scalar field: $\phi$} \\
\text{fermionic field: $\psi$} \,.
\end{array}
\ee
Given these correspondences we can now state the dynamical relationship between the
quanities involved: adding a source $J$ for an operator ${\mathcal{O}}$ in the QFT is
dual to imposing a boundary condition at infinity for the field $\phi$ corresponding to
${\mathcal{O}}$. The boundary condition is that as $r \to 0$ in the spacetime (\ref{SAH:eq:ads}), 
then the field $\phi$ tends towards the value $\delta \phi_{(0)} = J$, up to an overall power of $r$.
That is:
\be
\label{SAH:eq:backbone}
Z_\text{bulk}[\phi \to \delta \phi_{(0)}] = \left\langle \exp \left( i
\int d^dx \delta \phi_{(0)} {\mathcal{O}}\right) \right\rangle_\text{QFT} \,.
\ee
This formula is illustrated in figure \ref{SAH:response}. As anticipated in the figure,
the usefulness of expression (\ref{SAH:eq:backbone}) is that it will
enable us to compute expectation values and $n$ point functions of the operator ${\mathcal{O}}$
in the QFT by differentiating the left hand side with respect to $J=\delta \phi_{(0)}$.

From the basic relation (\ref{SAH:eq:backbone}), one can quickly obtain the following two important statements. See for instance \cite{SAH:Hartnoll:2009sz, SAH:McGreevy:2009xe} for a more leisurely exposition of these facts as well as the field-operator correspondence more generally. The first statement is that
\be
\begin{array}{c}
\text{Global symmetry (QFT)}
\end{array}
\quad \leftrightsquigarrow \quad
\begin{array}{c}
\text{Gauged symmetry (gravity)} \,.
\end{array}
\ee
This was implicit in our previous claim in (\ref{SAH:eq:list}) that a bulk Maxwell field was dual to a global current operator.
It is also closely related to the notion that gauge symmetries include `large' gauge transformations that can act on the boundary of spacetime as global symmetry operations.

Secondly, suppose that the bulk bosonic or fermionic fields are simply free fields with actions
\be\label{SAH:eq:scalaract}
S[\phi] = \int d^{d+1}x \sqrt{-g} \Big(  |\partial \phi - i q A \phi|^2 + m^2 |\phi|^2 \Big) \,, 
\ee
and
\be\label{SAH:eq:fermionact}
S[\psi] = \int d^{d+1}x \sqrt{-g} \left( i \bar \psi \Gamma \cdot \left(\partial + {\textstyle \frac{1}{4}} \omega_{\mu\nu} \Gamma^{\mu\nu} - i q A \right) \psi + m \bar \psi \psi  \right) \,,
\ee
respectively. For future convenience we have allowed the fields to have charge $q$ under some $U(1)$ gauge symmetry. Here the $\Gamma$s are Dirac gamma matrices and $\omega_{\mu\nu}$ is the spin connection
needed to describe fermions in a curved spacetime background. Given the above two actions, then the bosonic and fermionic operators ${\mathcal{O}}_B$ and ${\mathcal{O}}_F$, dual to $\phi$ and $\psi$ respectively, will have scaling dimensions
\be\label{SAH:eq:delta}
\Delta_B (\Delta_B - d) = (m L)^2 \,, \qquad \Delta_F = \frac{d}{2} + L m \,.
\ee
Recall that in the $d$ dimensional field theory, an operator ${\mathcal{O}}$ will be relevant if $\Delta_{\mathcal{O}} < d$, marginal if $\Delta_{\mathcal{O}} = d$ and irrelevant otherwise. From (\ref{SAH:eq:delta}) we can see that very massive bulk fields (in units of the AdS radius $L$) correspond to highly irrelevant operators. The computations we will outline in this chapter will involve a handful of light fields in the bulk. This is therefore dual to a handful of low dimensional operators controlling the dynamics of the QFT. Note that masses in the bulk do not correspond to masses in the dual QFT, which is still scale invariant.

We can outline the use of (\ref{SAH:eq:backbone}) to calculate a two point function in the strongly coupled CFT. For simplicity we will compute the Euclidean two point function, in terms of the imaginary time $\tau = i t$. We wish to evaluate
\be\label{SAH:eq:twopt}
\langle {\mathcal{O}}(x) {\mathcal{O}}(y) \rangle = \frac{\delta^2}{\delta\, \delta\phi_{(0)}^2} Z_\text{bulk}[\phi \to \delta \phi_{(0)}]   = \frac{\delta^2}{\delta\, \delta\phi_{(0)}^2} e^{-S[\phi \to \delta \phi_{(0)}]}\,.
\ee
In the second equality we have used the bulk classical limit (\ref{SAH:eq:limit}); $S[\phi \to \delta \phi_{(0)}]$ is the bulk action evaluated on a solution to the equations of motion subject to the boundary condition $\phi \to \delta \phi_{(0)}$. This is all very much in the spirit of Hamilton-Jacobi theory. We therefore need to solve the bulk equations of motion with general boundary conditions.

Consider an operator ${\mathcal{O}}$ dual to a neutral scalar field $\phi$ in the bulk with action (\ref{SAH:eq:scalaract}). The solution to the bulk equations of motion that is regular at the `AdS horizon' $r=\infty$ is given by the modified Bessel function
\be\label{SAH:eq:phisol}
\phi \propto \delta \phi_{(0)} r^{d/2} K_{\Delta_+ - \frac{d}{2}} \Big(r \sqrt{k^2 + \omega^2} \Big) e^{- i \omega \tau + i k \cdot x} \,.
\ee
We expressed the solution as a Fourier mode. The weight $\Delta_+$ is the larger of the two solutions to (\ref{SAH:eq:delta}). We now need to evaluate the action (\ref{SAH:eq:scalaract}) on the solution (\ref{SAH:eq:phisol}) and use the prescription (\ref{SAH:eq:twopt}).
This is complicated by the need to renormalise infinities coming from the infinite volume of AdS space. For a detailed discussion see e.g. \cite{SAH:Hartnoll:2009sz} and references therein; under the holographic correspondence these divergences are dual to the standard UV divergences encountered in quantum field theory. The upshot is a general result that applies to all types of operators and to Lorentzian as well as Euclidean signature. Near the boundary, the bulk solution will behave like
\be\label{SAH:eq:asymptotics}
\phi = \phi_{(0)} r^{d - \Delta} + \cdots + \phi_{(1)} r^\Delta + \cdots \,, \qquad (\text{as $r \to 0$}) \,.
\ee
This behaviour can be taken to define $\Delta$. The (retarded in the case of Lorentzian signature) two point function is then found to be
\be\label{SAH:eq:general2pt}
\langle {\mathcal{O}}(x) {\mathcal{O}}(y) \rangle_R = \frac{2 \Delta - d}{L} \frac{\phi_{(1)}}{\phi_{(0)}} \,.
\ee
The above formula is normalised by assuming that the bulk field $\phi$ has kinetic term $\frac{1}{2} (\partial \phi)^2$.
If this is not the case, the bulk field should be appropriately rescaled to put the kinetic term in this form. The rescaling will result in an extra overall factor in (\ref{SAH:eq:general2pt}).
Taking the $r \to 0$ limit of the solution (\ref{SAH:eq:phisol}) one easily obtains
\be\label{SAH:eq:OOresult}
\langle {\mathcal{O}} {\mathcal{O}} \rangle(k,\omega) \propto \Big(k^2 + \omega^2 \Big)^{-\frac{d}{2} + \Delta_+} \,.
\ee
This is the correct result for the Euclidean two point function of a scalar operator of scaling dimension $\Delta = \Delta_+$ in a $d$ dimensional CFT.

The result (\ref{SAH:eq:OOresult}) is completely fixed by conformal invariance. This would not be the case for higher point functions. The holographic correspondence will come into its own, however, when we place the QFT at a finite temperature.

\subsection{Finite temperature}

\begin{figure}[h]
\begin{center}
\includegraphics[height=120pt]{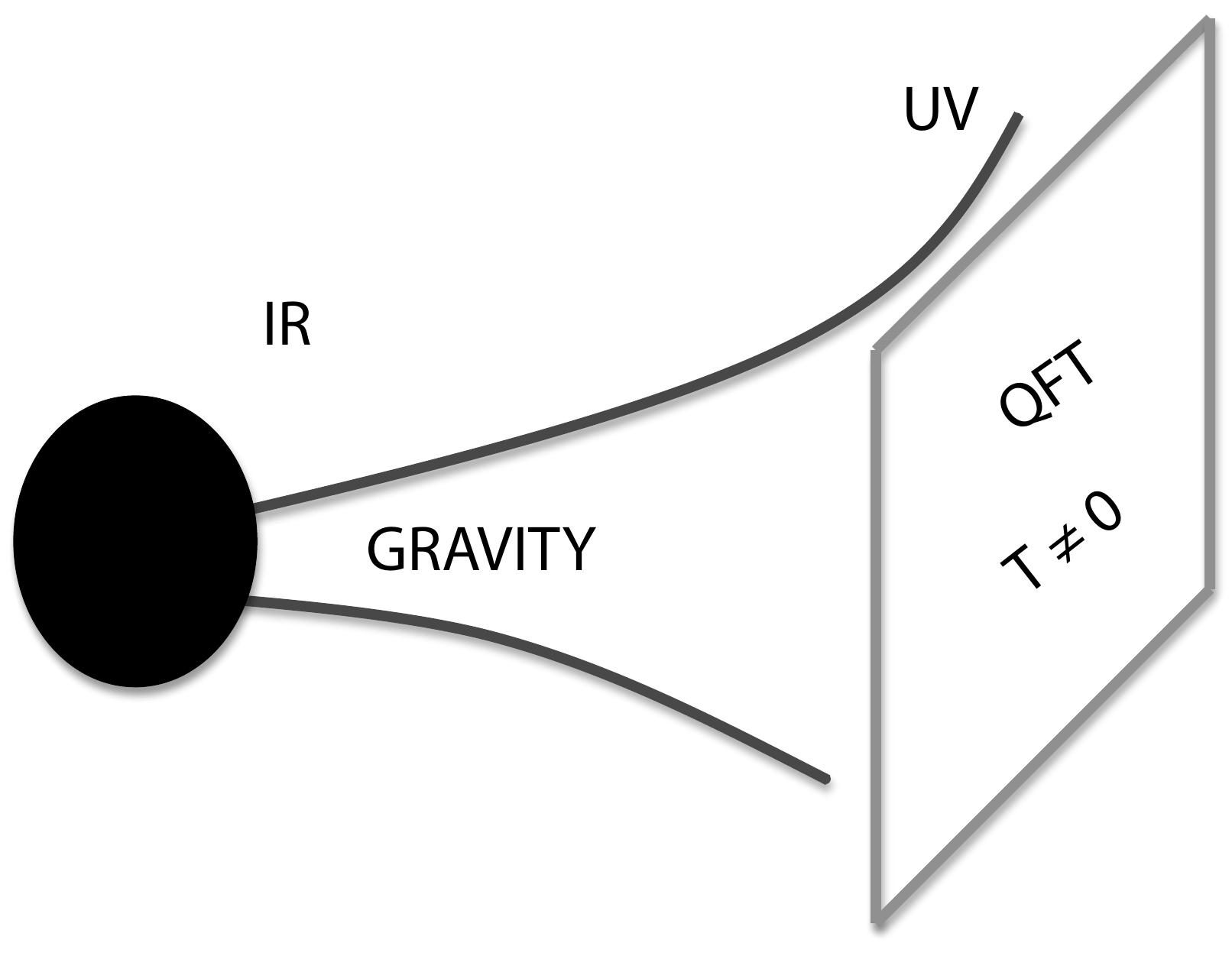}
\end{center}
\caption[Black hole and finite temperature.]{The strongly coupled QFT at finite temperature
is dual to classical gravity in a black hole spacetime.}\label{SAH:blackhole}
\end{figure}

Real time processes in a QFT at finite temperature can be difficult to compute, even at weak coupling.
A powerful feature of the holographic correspondence is that the finite temperature computations
are essentially no harder than computations at zero temperature: they still involve only
classical fields in a background curved spacetime. The important difference is that the spacetime is no
longer AdS space (\ref{SAH:eq:ads}) but rather an asymptotically AdS black hole, illustrated in figure
\ref{SAH:blackhole}.

While black holes may seem exotic, their appearance here is both natural and in
fact necessary. Given that all quantum field theories can be placed at a finite temperature,
the dual description of heating up the theory should not require any new ingredients to
be introduced. Thus we should still be able to use the action (\ref{SAH:eq:einstein}).
Temperature introduces an energy scale $T$ which must break the
scaling invariance (\ref{SAH:eq:scaling}). If we relax the scaling symmetry and look
for solutions to the Einstein equations (\ref{SAH:eq:eeqs}) which are still invariant under spatial rotations
and spacetime translations, it turns out that there is a unique regular solution:
the Schwarzschild-AdS black hole. This has metric
\be\label{SAH:eq:schwarzads}
ds^2 = \frac{L^2}{r^2} \left(- f(r) dt^2 + \frac{dr^2}{f(r)} + dx^i dx^i \right) \,,
\ee
where
\be\label{SAH:eq:fsads}
f(r) = 1 - \left(\frac{r}{r_+} \right)^{d} \,.
\ee
This solution is called a black hole because $f(r_+) = 0$, indicating
that light emitted to an asymptotic observer becomes infinitely redshifted
at the `event horizon' $r=r_+$.

The lengthscale $r_+$ determines the temperature of the dual field theory
via
\be\label{SAH:eq:T}
T = \frac{d}{4 \pi r_+} \,.
\ee
This is in fact the Hawking temperature of the black hole. Hawking radiation
however is a quantum mechanical effect in the bulk \cite{SAH:Hawking:1974rv}
and therefore suppressed in the large $N$ limit. A slightly formal argument that leads to
(\ref{SAH:eq:T}) can be made by considering the analytic continuation of the black
hole (\ref{SAH:eq:schwarzads}) to Euclidean signature $\tau = i t$ \cite{SAH:Gibbons:1976ue}. At the radius
$r=r_+$ the Euclidean time direction shrinks to a point. This is precisely like
the origin of polar coordinates: $dR^2 + R^2 d\theta^2$.
The only way such shrinking can occur without introducing a conical singularity is if
$\tau$ is periodic with period
\be\label{SAH:eq:periodic}
\tau \sim \tau + \frac{4 \pi}{| f'(r_+)|} = \tau + \frac{4 \pi r_+}{d} \,.
\ee
This period is obtained by changing coordinates so that the $\{r,\tau\}$ part of the
black hole metric, to leading order at $r \to r_+$, look like the $\{R,\theta\}$ metric for
flat space in polar coordinates. Requiring $\theta$ to have period $2\pi$ leads to
(\ref{SAH:eq:periodic}). The standard interpretation of the period of the Euclidean time circle
as the inverse temperature then gives (\ref{SAH:eq:T}).

The simplest finite temperature quantity we can calculate is the free energy of the theory.
This is given by
\be\label{SAH:eq:free}
F = - T \log Z = T S_E[g_\text{BH}] = - \frac{(4\pi)^d L^{d-1}}{2 \kappa^2 d^d} V_{d-1} T^d \,,
\ee
where in the second equality we have used the large $N$ relation $Z = e^{-S_E[g]}$, expressing the
partition function as the (Wick rotated, Euclidean) classical action evaluated on the black hole saddle point
(\ref{SAH:eq:schwarzads}). The final equality involves performing the evaluation. This again
requires renormalising divergences due to the infinite volume of the spacetime, details can
be found in \cite{SAH:Hartnoll:2009sz} and references therein. The dependence of the
free energy (\ref{SAH:eq:free}) on the spatial volume and temperature is fixed by the scale invariance
of the underlying QFT. The coefficient however gives physical information and is 
one definition of the central charge of the theory. Thus (\ref{SAH:eq:free}) 
confirms our previous statement that $F \sim c \,T^d$ with $c \gg 1$ given by (\ref{SAH:eq:limit}).
From the free energy (\ref{SAH:eq:free}) we can easily obtain the entropy
\be\label{SAH:eq:entropy}
S = - \frac{\partial F}{\partial T} = \frac{(4\pi)^{d} L^{d-1}}{2 \kappa^2 d^{d-1}} V_{d-1} T^{d-1}\,.
\ee

The black hole solution (\ref{SAH:eq:schwarzads}) describes the theory in equilibrium at finite
temperature, with the free energy (\ref{SAH:eq:free}). To perturb the system away from equilibrium
we simply need to perturb the black hole solution and compute response functions in the same
way as we did in the previous section at zero temperature. Dissipation due to the finite temperature is described
by matter falling through the black hole horizon.

\subsection{Spectral functions and quasinormal modes}

Retarded Green's functions at finite temperature are obtained via the holographic correspondence
in the same way as at zero temperature: One solves the bulk equations of motion of the
field $\phi$ dual to the operator ${\mathcal{O}}$ of interest. The equations are linearised about the
black hole background (\ref{SAH:eq:schwarzads}). Near the asymptotic boundary ($r \to 0$) the black
hole metric behaves in the same way as pure AdS space, and therefore the asymptotic solution for the field
$\phi$ will again take the form (\ref{SAH:eq:asymptotics}). The retarded Green's function is given by
(\ref{SAH:eq:general2pt}).

Generally wave equations in black hole backgrounds cannot be solved analytically. After
a Fourier decomposition
\be\label{SAH:eq:fourier}
\phi = \phi(r) \, e^{-i \omega t + i k \cdot x} \,,
\ee
one is left with an ODE for $\phi(r)$. It is straightforward however to solve the differential equation numerically and extract
the coefficients $\phi_{(0)}$ and $\phi_{(1)}$ from (\ref{SAH:eq:asymptotics}). In solving the differential equation
one needs to impose boundary conditions at the horizon $r=r_+$. The correct boundary conditions are so-called `ingoing' boundary conditions, in which bulk matter is falling into the black hole rather than coming out. This amounts to imposing
\be\label{SAH:eq:nonzeroT}
\phi(r) \propto e^{- i 4 \pi \omega/T \, \log(r-r_+)}  +  \cdots \qquad (\text{as} \quad r \to r_+) \,.
\ee
The key element here is the sign of the exponent. Combining (\ref{SAH:eq:nonzeroT}) and (\ref{SAH:eq:fourier}) we see that the wave moves towards the horizon $r=r_+$ as time advances. The temperature $T$ is given by (\ref{SAH:eq:T}).

The strongly coupled theories we are studying typically do not have quasiparticle excitations. Therefore, except for specific interesting circumstances to be addressed shortly, we do not expect sharp features in the response functions along the real frequency axis. This raises the question of
how best to characterise the spectral functions in such theories.

In all the known retarded Green's functions obtained
via the holographic correspondence, the only nonanalyticities found at finite temperature are poles at specific complex frequencies. At zero temperature the poles can degenerate into branch cuts, but at finite temperature the Green's functions appear to be meromorphic in the complex frequency plane.\footnote{Strictly speaking, we are referring to the analytic continuation of the retarded Green's function from the upper imaginary axis.} These poles are called the quasinormal frequencies of the Green's function, $\omega_\star$, and they largely determine the interesting structure of the function
\be
\langle {\mathcal{O}} {\mathcal{O}} \rangle_R (\omega) \sim \sum_{\omega_\star} \frac{c_\star}{\omega - \omega_\star} \,,
\ee
up to an overall function of $\omega$ that is holomorphic everywhere except possibly at infinity. It is clear from (\ref{SAH:eq:general2pt}) that the poles occur whenever
\be\label{SAH:eq:dirichlet}
\phi_{(0)}(\omega_\star) = 0 \,.
\ee
From the bulk point of view a quasinormal mode is a solution to the linearised equations of motion satisfying ingoing boundary conditions (\ref{SAH:eq:nonzeroT}) at the horizon together with (\ref{SAH:eq:dirichlet}) at the boundary \cite{SAH:Son:2002sd}. The quasinormal modes can therefore be determined numerically from the linearised bulk equations of motion using shooting techniques. We will see some examples below.

The quasinormal frequencies have an interesting structure -- as we will see in the following sections. For the moment we can note one circumstance in which quasinormal poles can approach the real frequency axis even at strong coupling. This is in the hydrodynamic regime. Hydrodynamic modes can be associated to conserved quantities or to broken symmetries and describe the long wavelength dynamics of the system (e.g. \cite{SAH:forster}).

Focus for a moment on the case of $d=3+1$ dimensional field theories. All
such theories will have hydrodynamic shear and sound modes, as these follow from
conservation of the energy momentum tensor. The shear mode, for example, is a quasinormal mode
of the form
\be\label{SAH:eq:shear}
\omega_\star = - i \frac{\eta}{\epsilon + P} k^2 \,, \qquad (\text{as $k \to 0$}) \,,
\ee 
where $\eta$ is the shear viscosity, $\epsilon$ the energy density and $P$ the pressure. Via the holographic correspondence the energy and pressure are determined from the free energy (\ref{SAH:eq:free}). The pole
can be found by computing the retarded Green's function of the $xy$ components of the energy-momentum
tensor: $\langle T^{xy} T^{xy}\rangle_R$ (taking the momentum to be along the $z$ direction).

As stated in (\ref{SAH:eq:list}), the operator dual to $T^{xy}$ is a fluctuation of the $g_{xy}$ metric component in the bulk.
We can therefore sketch a computation of the Green's function $\langle T^{xy} T^{xy}\rangle_R$. Firstly, we linearise the equations of motion (\ref{SAH:eq:eeqs}) about the black hole spacetime (\ref{SAH:eq:schwarzads}):
\be
ds^2 = ds^2_\text{BH} + 2 \delta g_{xy}(r) e^{- i \omega t + i k z} dx dy \,.
\ee
The resulting equation is
\be
r f (r f \delta g_{xy}')' = \left(8 f + k^2 r^2 f - \omega^2 r^2 - 4 f^2 \right) \delta g_{xy} \,.
\ee
To extract the Green's function, one now simply imposes ingoing boundary conditions at the black hole horizon (\ref{SAH:eq:nonzeroT}), solves the above equation (using say a small frequency and momentum expansion) and computes the Green's function from (\ref{SAH:eq:asymptotics}) and (\ref{SAH:eq:general2pt}). A shear mode pole (\ref{SAH:eq:shear}) is found with a shear viscosity $\eta$ most famously expressed as \cite{SAH:Kovtun:2004de}
\be\label{SAH:eq:etaovers}
\frac{\eta}{s} = \frac{1}{4 \pi} \,,
\ee
where the entropy density $s = S/V_3$ is given by (\ref{SAH:eq:entropy}). A very clear pedagogical exposition of this result can be found in \cite{SAH:Son:2007vk}. The result (\ref{SAH:eq:etaovers}) appears to be generic in the holographic correspondence to leading order at large $N$, but can have corrections with either sign (see \cite{SAH:Sinha:2009ev} for a summary of the status of the `viscosity bound conjecture' at the time of writing).

Other hydrodynamic coefficients, such as sound speeds and electrical conductivities can similarly be computed by exhibiting poles in Green's functions or by using Kubo formulae \cite{SAH:Son:2007vk}. For instance, the electrical conductivity $\sigma$ over the charge susceptibility $\chi$ is given by \cite{SAH:Kovtun:2008kx}.
\be
\frac{\sigma}{\chi} = \frac{1}{4 \pi T} \frac{d}{d-2} \,,
\ee
to leading order at large $N$.

Upon adding a chemical potential, and hence working with a finite charge density, we will see that there are two additional circumstances in which quasinormal modes can come close to the real frequency axis. The first is the case of bosonic operators, at the onset of superconducting instabilities, while the second is for fermionic operators, leading to strong coupling analogues of Fermi surfaces.

Before closing this section, we can note that beyond linear response, holographic techniques have recently
started to be applied to the study of far from equilibrium dynamics of strongly interacting theories \cite{SAH:Chesler:2008hg}.

\section{Finite chemical potential}

We started with the observation that at zero temperature the holographic correspondence gives a geometrical realisation of conformal invariance that allows computation of correlators in certain strongly coupled quantum critical theories. We then discussed the most universal deformation away from scale invariance, by placing the theories at a finite temperature. Another very important deformation away from scale invariance, especially in condensed matter systems, is to place the theory at finite chemical potential and hence induce a charge density.

Finite charge density means an expectation value for the time component of a global current $\langle J^t \rangle$. We claimed in (\ref{SAH:eq:list}) that the bulk dual to a global current is a Maxwell field. We therefore need to supplement our existing actions with the term
\be\label{SAH:eq:maxwell}
S[A] = - \frac{1}{4 g^2}  \int d^{d+1}x \sqrt{-g} F^2\,,
\ee
where $F=dA$ is the field strength. As with the Einstein-Hilbert action (\ref{SAH:eq:einstein}) there can in principle be higher derivative corrections to this action. Under a naturalness assumption, i.e. in the absence of very large dimensionless numbers, they will be suppressed in the large $N$ limit. This is known to occur in specific examples.

\begin{figure}[h]
\begin{center}
\includegraphics[height=140pt]{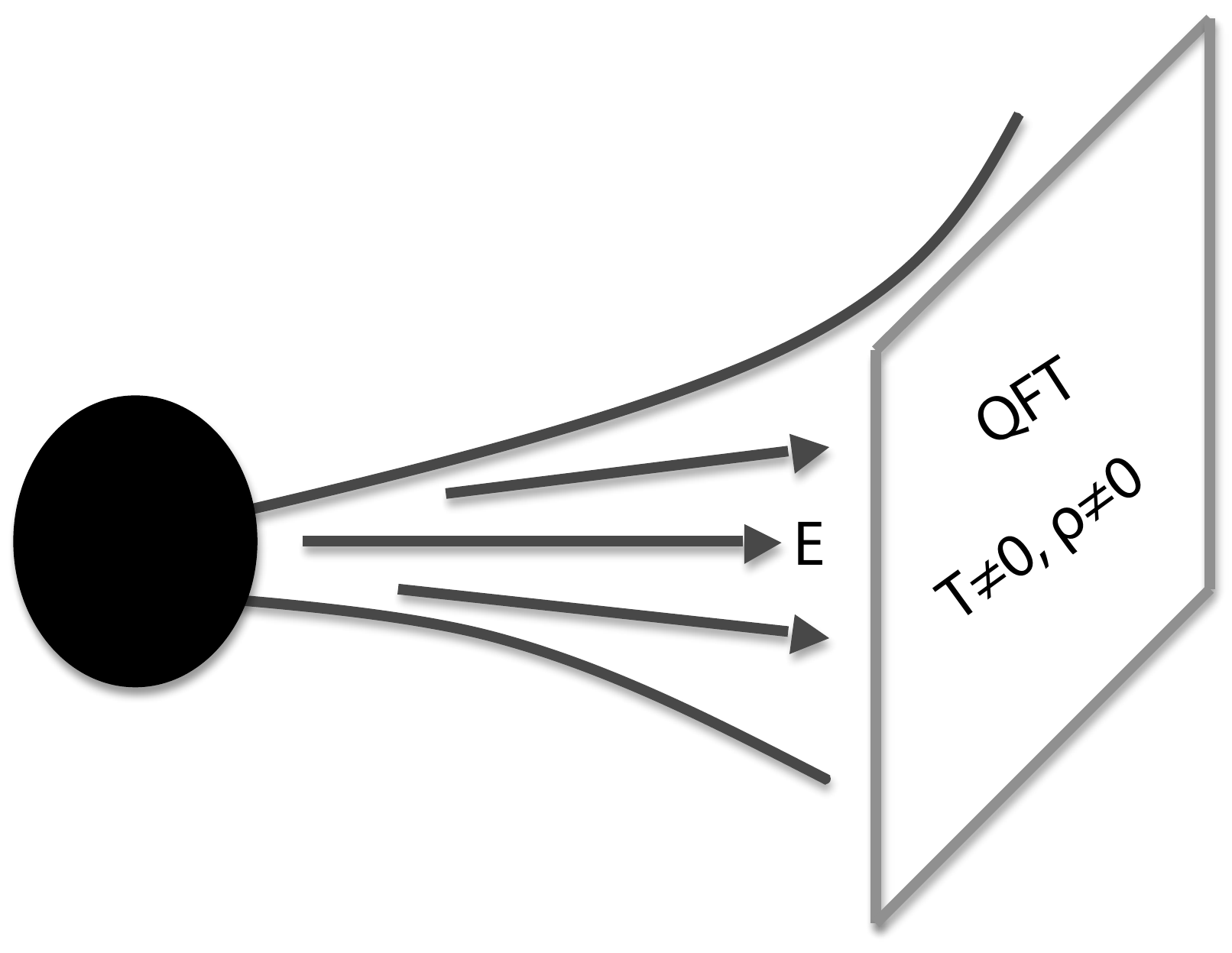}
\end{center}
\caption[Charged black hole and finite density.]{The strongly coupled QFT at finite temperature
and finite charge density is dual to classical gravity in a charged black hole spacetime.}\label{SAH:blackholeC}
\end{figure}

The gravitational dual description of working with a finite charge density in the field theory is to add an electric charge to the black hole. This is illustrated in figure \ref{SAH:blackholeC}.

There is a unique solution to Einstein-Maxwell theory, i.e. the theory given by the actions (\ref{SAH:eq:einstein}) plus (\ref{SAH:eq:maxwell}), that describes a black hole carrying an electric charge. This is called the Reissner-Nordstrom-AdS black hole. For a slower exposition of the following material see e.g. \cite{SAH:Hartnoll:2009sz, SAH:Herzog:2009xv}. The metric is given by
\be\label{SAH:eq:RNads}
ds^2 = \frac{L^2}{r^2} \left(- f(r) dt^2 + \frac{dr^2}{f(r)} + dx^i dx^i \right) \,,
\ee
where now
\be\label{SAH:eq:RNf}
f(r) = 1 - \left(1 + \frac{r_+^2 \mu^2}{\gamma^2} \right) \left(\frac{r}{r_+}\right)^d +
\frac{r_+^2 \mu^2}{\gamma^2} \left(\frac{r}{r_+}\right)^{2(d-1)} \,.
\ee
In this last expression we defined
\be\label{SAH:eq:gamma}
\gamma^2 = \frac{(d-1) g^2 L^2}{(d-2) \kappa^2} \,,
\ee
which is a dimensionless measure of the relative strengths of the gravitational
and Maxwell forces. The nonzero Maxwell potential is
\be\label{SAH:eq:At}
A_t = \mu \left[1 - \left(\frac{r}{r_+}\right)^{d-2} \right] \,.
\ee

The above solution is parametrised by two scales: the chemical potential of the field theory $\mu$
and the horizon radius $r_+$. The latter is related to the temperature by
\be\label{SAH:eq:RNT}
T = \frac{1}{4 \pi r_+} \left(d -  \frac{(d-2) r_+^2 \mu^2}{\gamma^2} \right) \,.
\ee
As in the neutral case above, the temperature is determined by regularity of the Wick rotated
Euclidean spacetime at the horizon $r=r_+$. By evaluating the action on shell one can
obtain the free energy
\be\label{SAH:eq:Free2}
\Omega(T,\mu) = - \frac{L^{d-1}}{2 \kappa^2 r_+^d} \left(1 +  \frac{ r_+^2 \mu^2}{\gamma^2}  \right) V_{d-1} = {\mathcal{F}}\left(\frac{T}{\mu}\right) V_{d-1} T^d\,,
\ee
where in the second equality $r_+$ has been eliminated in favour of the temperature by using (\ref{SAH:eq:RNT}). By scale invariance of the underlying quantum critical theory, the only nontrivial dependence on temperature and chemical
potential must be in the dimensionless ratio $T/\mu$.

From the free energy (\ref{SAH:eq:Free2}) one can again determine the entropy $S = - \partial_T \Omega$, as well as the charge density $\rho V_d  = - \partial_\mu \Omega$ and other functions of state. One exotic feature that is found
is that the entropy remains finite at zero temperature. At the time of writing it is not clear whether this constitutes a prediction of the holographic correspondence (`critical spin ice'?), an artifact of the large $N$ limit, or perhaps an indication that the Reissner-Nordstrom black hole spacetime is not capturing the correct finite density ground state of the theory.

Taking the above black hole as the dual of the theory at finite density and temperature, two natural questions arising as the system is cooled down are firstly whether it develops superconductivity and secondly whether there are indications of a sharp Fermi surface. At weak coupling the answers to these questions would only depend on the existence of charged bosonic and/or fermionic quasiparticles. The strongly coupled picture that emerges from the holographic correspondence is somewhat more subtle.

\subsection{Bosonic response and superconductivity}

\begin{figure}[h]
\begin{center}
\includegraphics[height=140pt]{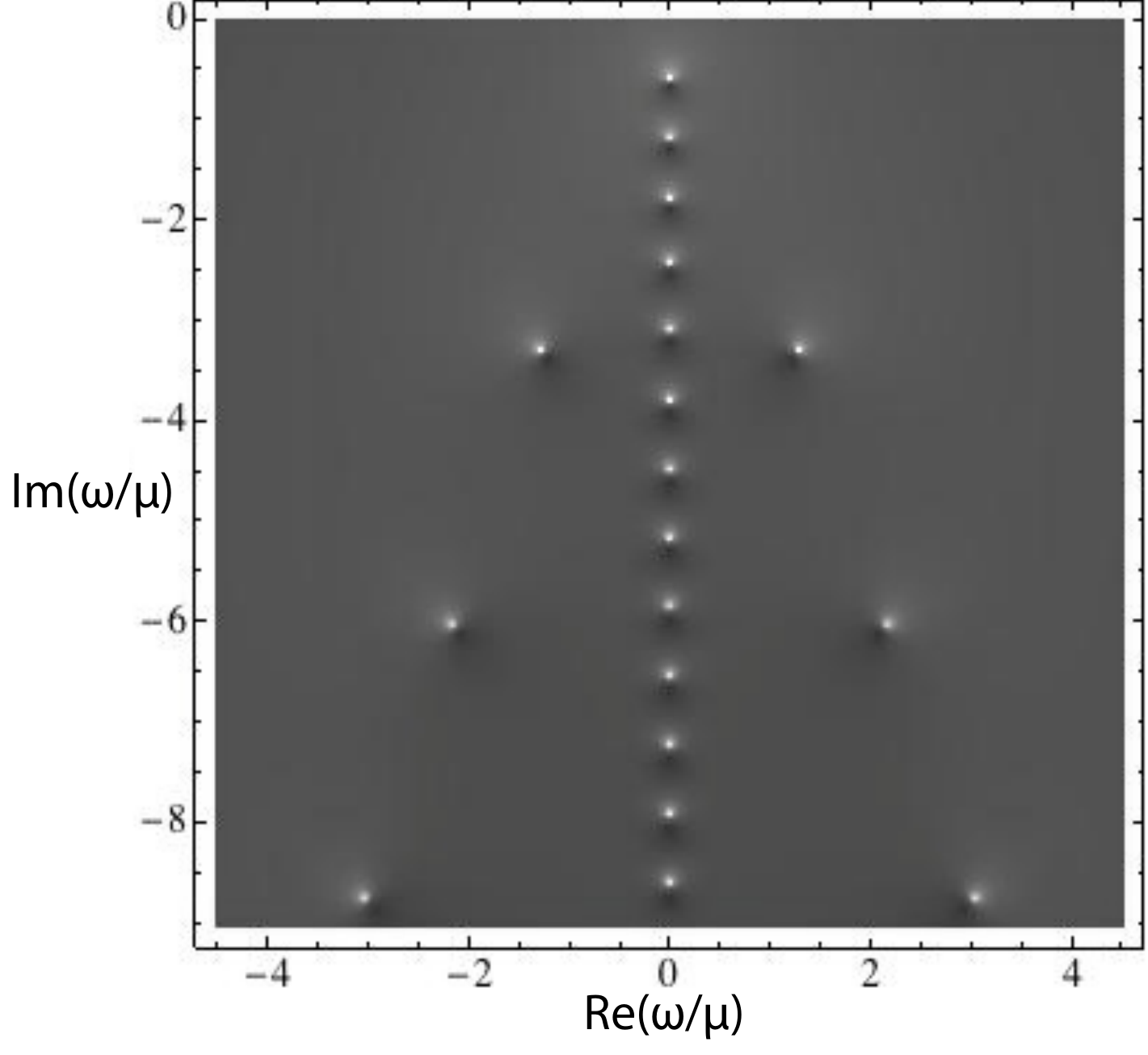}
\includegraphics[height=140pt]{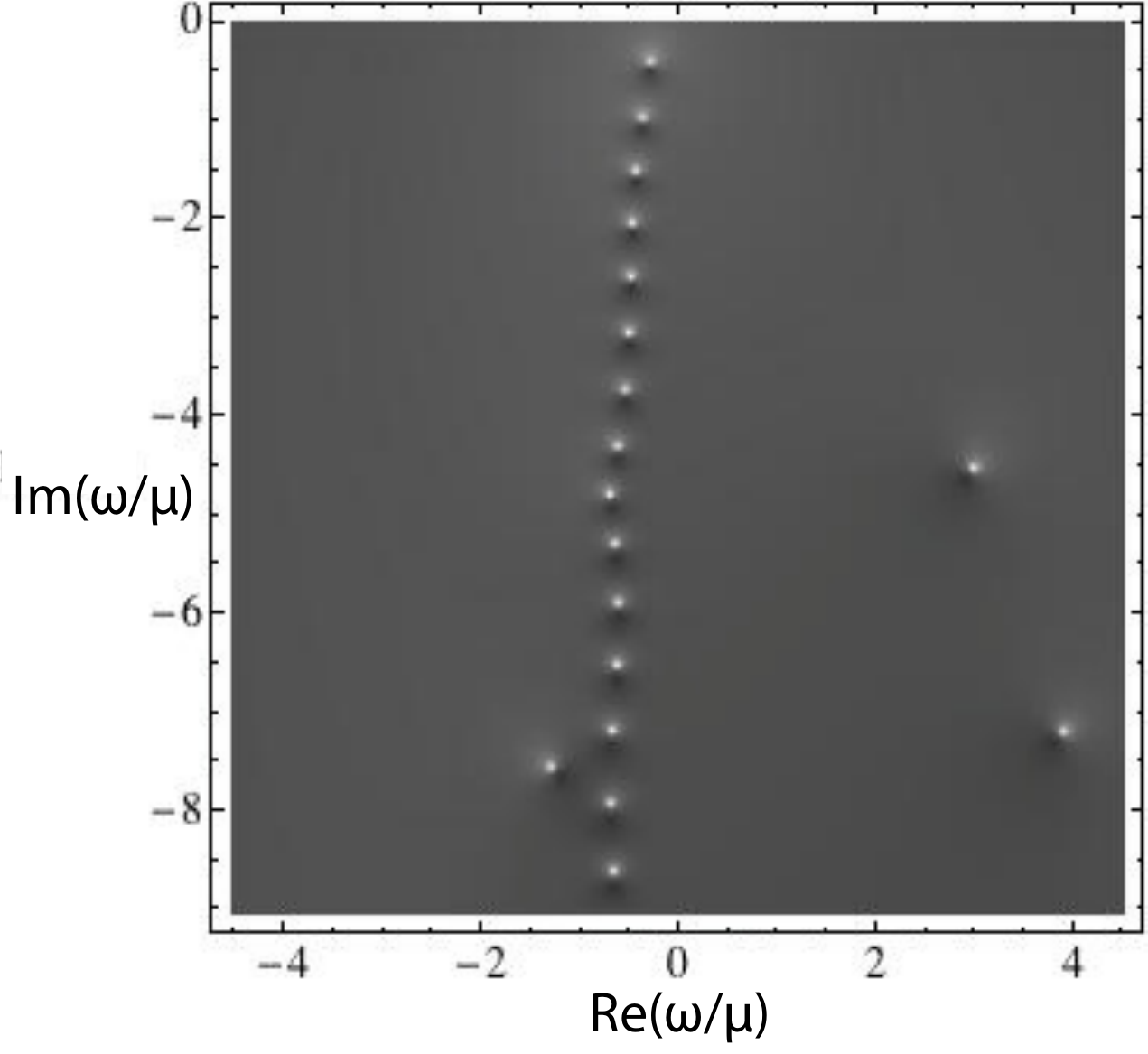}
\end{center}
\caption[Quasinormal poles of charged bosons.]{Zero momentum quasinormal poles of charged bosons at temperature $T/\mu = 0.075$. The operators ${\mathcal{O}}$ have scaling dimension $\Delta=3$ in 2+1 dimensions and charges $q=0$ (left plot) and $q=2$ (right plot).}\label{SAH:quasinormal}
\end{figure}

At this point we know nothing about what exactly is carrying the charge density.
In this section we probe the finite density state with bosonic operators. The following section will address
fermionic probes. Suppose that our strongly coupled field theory has at least one charged bosonic operator ${\mathcal{O}}$. This will be dual to a charged scalar field $\phi$ in the bulk with an action along the lines of (\ref{SAH:eq:scalaract}). We can compute the spectral density $ \text{Im} \, \langle {\mathcal{O}}{\mathcal{O}}\rangle_R(\omega,k)$ using the methods discussed above. As we have suggested, an instructive way to think about the spectral density at strong coupling is to compute the quasinormal frequencies of the Green's function. Some illustrative results are shown in figure \ref{SAH:quasinormal}. For details of the computation see \cite{SAH:Denef:2009yy}.

The quasinormal poles of a neutral operator, left hand plot in figure  \ref{SAH:quasinormal}, come in two types spaced by either the temperature or the chemical potential scales. As the charge of the operator is increased, right hand plot, the `chemical potential' poles move to the right and can merge with the `temperature' poles. The physics of these mergers has not yet been elucidated.

The other important pole motion in figure  \ref{SAH:quasinormal} is that as the charge of the operator is increased, the topmost pole moves closer to the real axis. If the charge is increased sufficiently, the pole will move into the upper half frequency plane, indicating an exponentially growing mode and an instability of the vacuum. When this occurs, the operator  ${\mathcal{O}}$ will gain an expectation value, leading to a superconducting state \cite{SAH:Gubser:2008px, SAH:Hartnoll:2008vx, SAH:Hartnoll:2008kx}. For scalar fields with action (\ref{SAH:eq:scalaract}) there is a precise criterion for when an operator ${\mathcal{O}}$ is unstable against condensing. In $d=2+1$ this occurs at some temperature $T_c$  if \cite{SAH:Denef:2009tp}
\be\label{SAH:eq:criterion}
q^2 \gamma^2 \geq 3 + 2 \Delta_{\mathcal{O}} (\Delta_{\mathcal{O}} - 3) \,.
\ee
Recall that $\gamma$ was defined in (\ref{SAH:eq:gamma}), effectively setting the units of charge, while $q$ and $\Delta_{\mathcal{O}}$ are respectively the charge and scaling dimension of ${\mathcal{O}}$.

There are several important lessons from (\ref{SAH:eq:criterion}), which is a criterion for the onset of superconductivity in a strongly coupled field theory. There are no quasiparticles or `glue' or `pairing mechanism' in sight. Instead the instability is described in terms of the motion of quasinormal poles and the criterion for instability is phrased in terms of the scaling dimension and charge of operators in a quantum critical theory.

\begin{figure}[h]
\begin{center}
\includegraphics[height=115pt]{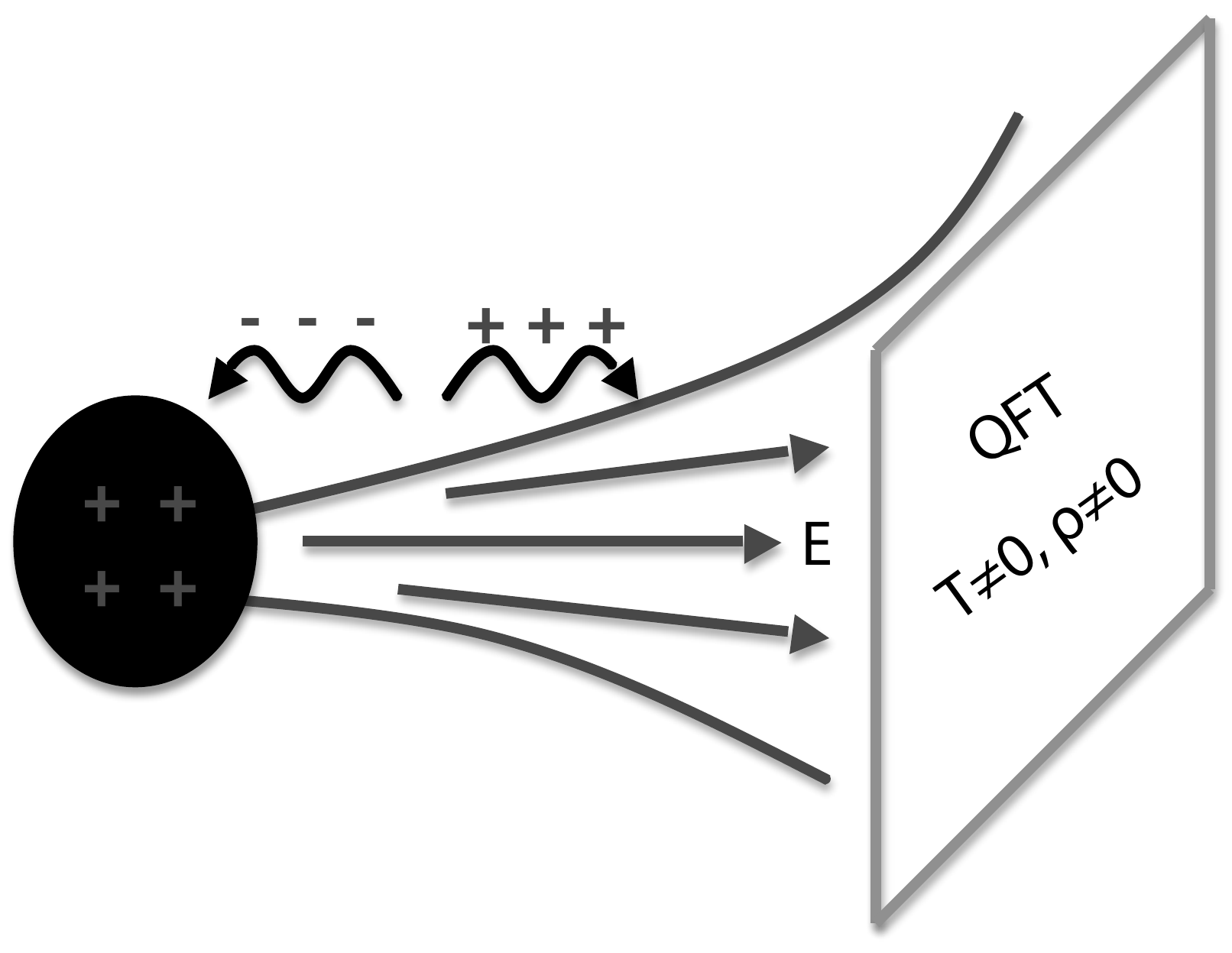}\hspace{0.5cm}
\includegraphics[height=115pt]{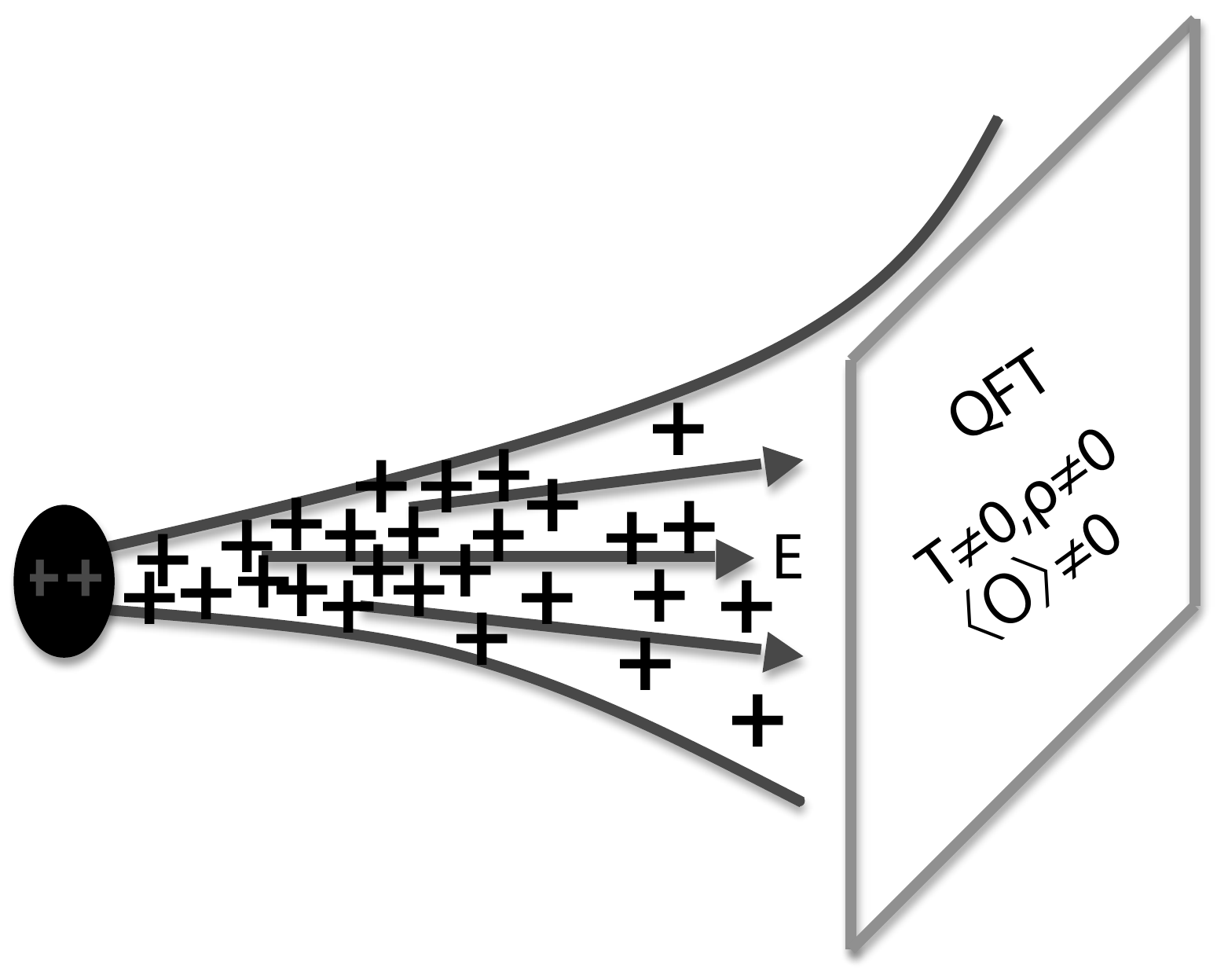}
\end{center}
\caption[Superconducting black hole instability.]{The superconducting instability of black
holes can be thought of as a polarisation of the spacetime. Below $T_c$ the charge is largely carried by the
condensate rather than the black hole. Such solutions are called hairy black holes.}\label{SAH:instability}
\end{figure}

The spacetime picture of the onset of superconductivity is shown in figure \ref{SAH:instability}. The instability is closely related to so-called `superradiance instabilities' of black holes. Below the instability temperature $T_c$, the Reissner-Nordstrom black hole (\ref{SAH:eq:RNads}) is no longer the correct ground state. Instead one must solve the coupled Einstein-Maxwell-scalar equations of motion to find new solutions in which the scalar $\phi$ is nonzero. These are called `hairy' black holes. For more details and references (for instance the computation of $T_c$, $\langle {\mathcal{O}} \rangle (T)$, and spectral functions in the superconducting state) see e.g. \cite{SAH:Hartnoll:2009sz, SAH:Herzog:2009xv}.

\subsection{Fermionic response and non-Fermi liquids}

While many theories with holographic duals do indeed develop superconducting instabilities at low temperatures \cite{SAH:Denef:2009tp}, it is not clear that they all do. It is an open question whether there necessarily exist
operators in the QFT satisfying (\ref{SAH:eq:criterion}). Interestingly this question is dually related to the `weak gravity conjecture' \cite{SAH:ArkaniHamed:2006dz, SAH:Denef:2009tp}. In cases where
the Reissner-Nordstrom black hole (\ref{SAH:eq:RNads}) is stable down to low temperature (or, alternatively, removing the superconducting phase with a large magnetic field) then one can study the low temperature fermionic response, in search for signs of a Fermi surface. This was first attempted in \cite{SAH:Lee:2008xf} while a comprehensive analytic understanding was achieved in \cite{SAH:Faulkner:2009wj}. We shall very briefly outline some of the ideas arising from \cite{SAH:Faulkner:2009wj}.

Consider a charged fermionic operator $\Psi$ dual to a bulk fermion with action (\ref{SAH:eq:fermionact}). The spectral density of this operator at zero temperature and finite charge density is obtained following the prescription described above. It was found in \cite{SAH:Faulkner:2009wj} that if the charge of the fermionic operator is sufficiently big compared to its scaling dimension (cf. the criterion for superconductivity in (\ref{SAH:eq:criterion})) then there is a quasinormal pole with the dispersion relation
\be\label{SAH:eq:dispersion}
\frac{\omega_\star}{v_F} + h e^{i \theta} \omega_\star^{2 \nu} = k - k_F \,,   
\ee
where $v_F, k_F, \nu, \theta$ and $h$ are positive constants. The pole bounces off the real axis as the momentum $k$ reaches the `Fermi momentum' $k_F$. This is characteristic of Fermi liquids. However, the residue of the pole is found to vanish as it hits the real axis, and therefore there is a not a well defined quasiparticle. Furthermore, the dispersion relation (\ref{SAH:eq:dispersion}) is not that of a Fermi liquid. 
In general, the low energy dispersion relation $\omega_\star(k)$, following from (\ref{SAH:eq:dispersion}), depends on whether $\nu$ is greater or less than $\frac{1}{2}$. The case $\nu = \frac{1}{2}$ is an interesting separate case and leads to
\be
\frac{\omega_\star}{v_F} + h e^{i \theta} \omega_\star \log \omega_\star = k - k_F \,,   
\ee
which is precisely of the form required for a `marginal Fermi liquid' originally postulated on phenomenological grounds
to describe the cuprates \cite{SAH:marginal}.

\begin{figure}[h]
\begin{center}
\includegraphics[height=130pt]{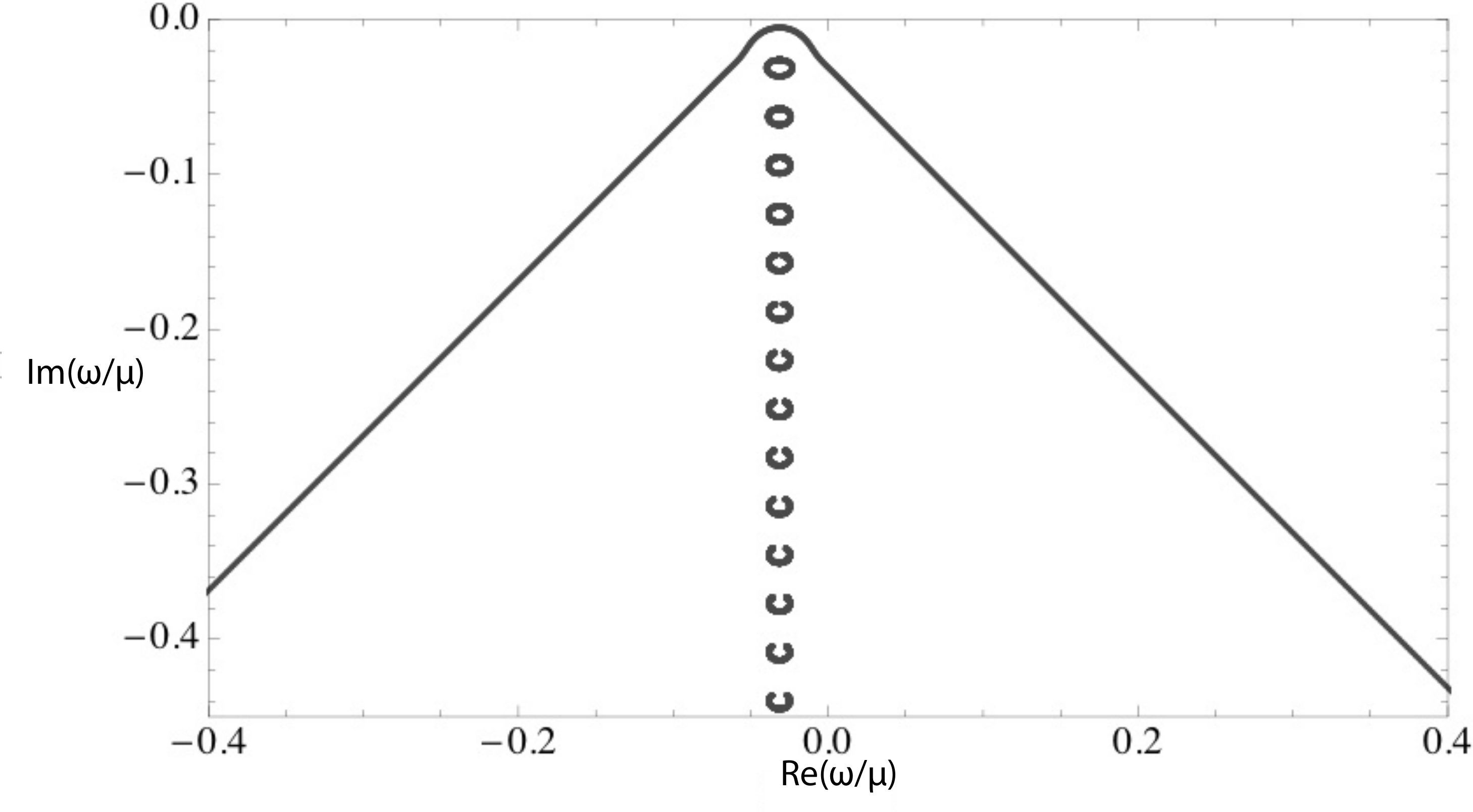}
\end{center}
\caption[Quasinormal poles and Fermi surfaces.]{Motion of quasinormal poles at low temperature as $k$ is varied through $k_F$.}\label{SAH:poledancing}
\end{figure}

The upshot of the previous paragraph is that the holographic correspondence at finite density and zero temperature leads to computationally controlled non-Fermi liquid behaviour of fermion spectral densities. No assumption about the existence of quasiparticles is made, instead the dispersion relation emerges from a quasinormal mode of a fermionic operator which has been consistently treated together with all the other quasinormal modes.  An illustrative motion of the quasinormal modes $\omega_\star(k)$ at a finite low temperature is shown in figure \ref{SAH:poledancing}. More details can be found in \cite{SAH:Denef:2009yy, SAH:Faulkner:2009wj}.

The poles close to the negative imaginary frequency axis in figure \ref{SAH:poledancing}  coalesce at zero temperature to form the branch cut visible in equation (\ref{SAH:eq:dispersion}). The same occurs with the bosonic poles in figure \ref{SAH:quasinormal}. It is clear in (\ref{SAH:eq:dispersion}) that this branch cut is intimately connected with the non-Fermi liquid dispersion of the quasinormal pole. In figure \ref{SAH:poledancing}  one sees that the branch cut poles also move significantly (in circles), as the pole of (\ref{SAH:eq:dispersion}) moves up to and bounces off the real frequency axis. To correctly compute the effect of `pole dancing' on physical quantities it will generally be important to consider all the poles close to the real axis, not just the pole with dispersion relation (\ref{SAH:eq:dispersion}). For an example see 
\cite{SAH:Denef:2009yy}. This is a crucial difference in general between quasinormal modes and quasiparticles. While we have a well defined non-Fermi liquid, with a dispersion relation (\ref{SAH:eq:dispersion}), strong coupling requires us to consider the dynamics of the branch cut as well as the `almost' quasiparticle pole. Unlike almost all conventional models of non-Fermi liquids, the holographic correspondence allows this to be done in a controlled way.

\section{Current and future directions}

At zeroth order, the holographic correspondence gives a geometrical perspective on quantum criticality that is particularly amenable to the study of finite temperature real time processes, such as response functions and also far from equilibrium dynamics. In a large $N$ limit, that nontheless retains the strongly coupled nature of the theory, computation of all of these quanitites is reduced to solving classical gravitational equations in one higher dimension than the field theory. The calculations are both significantly easier and conceptually more transparent than conventional approaches.

Current excitement about the possibilities of an applied holographic correspondence has been deepened, however, by recent discoveries concerning the strongly interacting theories at finite charge density and low temperatures. We have outlined above how charged bosonic operators were found to lead to superconducting phases while charged fermionic operators captured non-Fermi liquid behaviour. It is likely that deepening our understanding of these two phenomena, and the connection between them, will be an important focus of future research.

The onset of superconductivity in holographic models was found to be mediated by the charge and scaling dimension of operators in the quantum critical theory. These are the natural and well defined quantities to consider at strong coupling, rather than `electrons' and `glue' and `pairing'. It will be important to understand if criteria such as equation (\ref{SAH:eq:criterion}) can be applied generally to strongly interacting systems, without reference to a holographic description.

While field theories that become superconducting according to (\ref{SAH:eq:criterion}) are interesting, equally interesting are those that do not. At weak coupling, a theory with gapless charged bosonic excitations will necessarily develop condensates when placed at a nonzero chemical potential. This statement is not true at strong coupling if there are no operators in the theory satisfying (\ref{SAH:eq:criterion}). It will be interesting to explore if these theories can provide realisations of exotic `Boson metal' phases.

Non-Fermi liquid dispersion such as (\ref{SAH:eq:dispersion}) of poles in a fermion propagator that are not weakly interacting quasiparticles hints at a notion of a strongly coupled Fermi surface. This concept needs to be developed further. In \cite{SAH:Denef:2009yy} the magnetic susceptibility in these theories was computed as a function of the magnetic field, in a search for quantum oscillations that are characteristic of Fermi surfaces. At low temperatures, the anticipated periodic nonanalyticities were found, taking the schematic form
\be\label{SAH:eq:oscillations}
\chi = - \lim_{T \to 0} \frac{\partial^2\Omega}{\partial B^2} \sim + \sum_\ell \,  \Big| \ell B - \frac{A_F}{2 \pi q}\Big|^{-2+1/2\nu} \,,
\ee
where $A_F = \pi k_F^2$ is the Fermi surface area and $\nu$ is as in (\ref{SAH:eq:dispersion}). These divergences are softer than the usual weak coupling divergences in the susceptibility, which become delta functions at zero temperature. At the time of writing, quantum oscillations are proving to be valuable probes of the cuprate superconductors. It will be important to identify any generic qualitative differences between weak and strong coupling both in order to interpret data correctly and also to have signatures for novel states of matter.

Obtaining the result (\ref{SAH:eq:oscillations}) involved developing a formula for the leading correction in $1/N$ to the large $N$ result. It was shown that the correction to the free energy due to a specific bulk field could be expressed as a sum over the corresponding quasinormal frequencies
\be\label{SAH:eq:meisterfermions}
\Omega = \,  \Omega_0 +  T \sum_{\omega_\star} \log \left( \frac{1}{2 \pi}
\left| \Gamma \left(\frac{i \omega_\star}{2 \pi T} + \frac{1}{2} \right) \right|^2 \right)  + \cdots \,,
\ee
where $\Omega_0$ is the leading order large $N$ result (\ref{SAH:eq:Free2}) of the Reissner-Nordstrom black hole.
The above formula is for fermionic operators. Expressions like (\ref{SAH:eq:meisterfermions}) reinforce the idea that
quasinormal modes may be the correct and required generalisation of quasiparticles to describe strongly interacting systems. Indeed it seems likely that (\ref{SAH:eq:meisterfermions}) can be generalised beyond the holographic correspondence and to other observables such as response functions.

A possible fly in the ointment is that the non-Fermi liquid dispersion (\ref{SAH:eq:dispersion}) is closely tied to the presence of a branch cut in the zero temperature Green's function. This in turn is directly related to the `$AdS_2$' near horizon region of the zero temperature Reissner-Nordstrom black hole \cite{SAH:Faulkner:2009wj, SAH:Denef:2009yy}. Finally, this $AdS_2$ region is ultimately responsible for the finite entropy of the system at zero temperature. It remains to be seen whether interesting fermion dispersion relations can be obtained without a finite entropy at zero temperature.
The zero temperature limit of superconducting states appear to have a vanishing entropy, (see e.g. \cite{SAH:Gubser:2009cg, SAH:Horowitz:2009ij} for an extended discussion of the zero temperature limit, following earlier work); it will be very interesting to see whether these states show signs of Fermi surfaces or not. More generally the fate of the quasinormal poles (\ref{SAH:eq:dispersion}) across a superconducting transition is an important open question.

A somewhat distinct direction of future research will be the study of disorder and impurities in strongly interacting systems. Disorder plays a crucial role in many systems of technological and theoretical interest. The holographic correspondence will provide a framework in which familiar phenomena such as Anderson localisation or the Kondo effect can be studied at strong coupling. A first attempt to include weak disorder can be found in \cite{SAH:Hartnoll:2008hs}.
\vspace{0.5cm}

{\bf Acknowledgements:} The perspective taken in this chapter was developed in discussions with many people, in particular the participants in the KITP workshop `Quantum Criticality and the AdS/CFT 
Correspondence'. I am also grateful to Lars Fritz for helpful comments on the text. This text will become a chapter of the book ``Understanding Quantum Phase Transitions,'' edited by Lincoln D. Carr (Taylor \& Francis, Boca Raton, 2010).


\clearpage
\printindex

\end{document}